\documentclass[12pt,preprint]{aastex}

\shorttitle{An ionized Fe-K edge in APM 08279+5255}
\shortauthors{Hasinger et al.}

\begin{document}

\title{Discovery of an ionized Fe-K edge in the z=3.91 Broad 
    Absorption Line Quasar APM 08279+5255 with {\it XMM--Newton}}

\author{G. Hasinger\altaffilmark{1}, N. Schartel\altaffilmark{2}, and S. Komossa\altaffilmark{1}}
\affil{$^1$Max-Planck-Institut f\"ur extraterrestrische Physik,
Postfach 1312, D-85741 Garching, Germany}
\affil{$^2$XMM--Newton Science Operation Center, European Space Agency, Villafranca del Castillo, Apartado 50727, E-28080 Madrid, Spain}

\begin{abstract}
Recent
{\it XMM--Newton} observations of the high-redshift, lensed, broad absorption line (BAL) quasi-stellar object APM 08279+5255, 
one of the most luminous objects in the universe, 
allowed the detection of a high column density absorber ($N_H \approx 10^{23}$~cm$^{-2}$) 
in the form of a K-shell absorption edge of significantly ionized iron 
(Fe XV - XVIII) and corresponding ionized lower--energy absorption.  
Our findings confirm a basic prediction of phenomenological geometry
models for the BAL outflow and can constrain the size of the 
absorbing region. The Fe/O abundance of the 
absorbing material is significantly higher than solar (Fe/O = 2--5),
giving interesting constraints on the gas enrichment history in the 
early Universe.

\end{abstract}


\section{Introduction}
Absorption Line (BAL) Quasi-stellar Objects (QSOs) are a key to
understand
the geometry and physical state of the medium in the immediate vicinity of
accreting supermassive black holes. A new unified model
(Elvis 2000)  indicates that a significant fraction
of the matter accreted into the region of the compact object is flowing out
again. On either side of the accretion disk it should form a funnel-
shaped
shell, in which the outflowing gas is ionized and accelerated to velocities of
0.05-0.1$c$ by the powerful radiation force of the central object. If the
observer's inclination is favorable, this flow intercepts the line of sight
with a large column density and produces the blue-shifted broad UV absorption
line features observed in about 10\% of all QSOs. However, the UV/optical
spectra
sample only a minor fraction of the total column density of the flow, which is
predicted to be highly ionized so that it mainly absorbs X--rays.

APM 08279+5255 is an exceptionally luminous Broad Absorption Line (BAL) QSO at
redshift z=3.91 (Irwin et al., 1998), coincident with an IRAS Faint Source
Catalogue object and was also detected at 850 $\mu$m with SCUBA, implying an
apparent far-infrared luminosity of $> 5 \times 10^{15}~L_{\odot}$
(Lewis et al., 1998). The object is
strongly lensed (Ledoux et al., 1998; Ibata et al., 1999; Egami et al., 2000), 
with a magnification factor of 50-100, but even taking
this magnification factor into account, the object is still among the
most luminous in the Universe. The low resolution discovery spectrum 
of APM 08279+5255 showed a broad
absorption trough (BAL) on the blue side of Ly$_\alpha$.
An excellent high resolution optical spectrum was obtained with HIRES
at the Keck telescope (Ellison et al., 1999) and a detailed study of the
physical conditions in the broad
absorption line flow of the QSO (Srianand \& Petitjean, 2000)
came to the conclusion, that the
corresponding gas stream, outflowing with velocities up to 12000 km~s$^{-1}$,
is heavily structured and highly ionized.

In this paper we report X--ray observations of APM 08279+5255 
with {\it XMM--Newton} obtained in October 2001 and April 2002,
where we detected an ionized Fe K edge in the continuum 
of the QSO which is very likely related to the UV broad absorption 
line flow. In section 2 we present the X--ray observations and 
analysis. In section 3 we discuss the results. Throughout this 
paper we use a Hubble constant of  50~km~s$^{-1}$~Mpc$^{-1}$ and a deceleration
parameter $q_0=0.5$.  

\section{X--ray Observations and analysis}

The first
X--ray observation of APM 08279+5255 was taken  on October 11, 2000 with the
{\it Chandra} observatory (Gallagher et al., 2002). {\it XMM-Newton} observed
APM 08279+5255 for 15 ksec on October 30, 2001 
in good background conditions. The EPIC spectrum revealed  a tantalizing
absorption feature around 1.5 keV, close to the energy
expected from highly redshifted iron, either associated with the
object itself or with intervening material. Unfortunately the
statistical quality of the data did not allow more quantitative
analysis. A proposal for director's discretionary time for
a significantly longer observation (100 ksec) was accepted by the 
{\it XMM--Newton}
project scientist and executed on April 28--29, 2002. 
Table 1 gives details for the {\it XMM--Newton} and {\it Chandra} observations.

The XMM data were analyzed using the latest version of the standard analysis 
software {\it SAS 5.0}. 
The EPIC pn--CCD detector detected the source with an average count rate of 
$\sim 0.19$ cts~s$^{-1}$ and $\sim 0.15$ cts~s$^{-1}$, in the XMM1 and XMM2 observation, 
respectively. Within each of the observations there is no significant 
time variability detected.  
Fig. 1 shows the normalized pn--CCD counts spectrum of the source
from the 100 ksec XMM2 observation. The source is clearly detected out to 12
keV (almost 60 keV in the restframe of the quasar!). Using {\it xspec version
11.1.0}, we fitted the combined pn--CCD and MOS--CCD spectra and in all fits included
a neutral absorption with the Galactic $N_H$ value fixed to 
$4 \times 10^{20}$~cm$^{-2}$.
Following Gallagher et al. (2002), we first assumed a simple power law
spectrum with an intrinsic cold gas absorber in the rest frame of the QSO
(z=3.91).
The two {\it XMM--Newton} (EPIC pn+MOS) and one {\it Chandra} (ACIS-I) dataset 
can be fit with the same model and,
apart from a $\sim 20\%$ flux decrease between the first and the last
observation, the spectral fit parameters are consistent between the three
datasets. The best fit values for all three datasets are given in
Table 2.

The spectral fit residuals of the 100 ksec XMM2 observation,
however, show systematic deviations across the whole spectral range (Fig. 1).
The fit is
statistically unacceptable (reduced $\chi^2$ = 1.45 for 126 degrees of
freedom),
the most significant deviation is an edge-like feature around 1.55 keV,
corresponding to roughly 7.7 keV in the rest frame of the object. 
There is no evidence for an emission line associated
with the K edge: the upper limit for the equivalent width of a narrow 
emission line in the rest-frame energy range 6.4-6.9 keV is about 110 eV,
much lower than those found e.g. from ionized reflection disk models.
We therefore interpret
the observed spectral feature as an absorption edge due to ionized iron in the BAL flow
of APM 08279+5255. Indeed, if we include an additional absorption edge in the
spectral model we obtain a highly significant improvement of the fit
($\Delta \chi^2 = -87$ for two additional parameters) for the XMM2 dataset
with
an edge energy of $E_{edge} = 7.7 \pm 0.1$ keV and an optical depth of
$\tau = 0.46 \pm 0.07$. An improved fit is also obtained for the
XMM1 observation, with independently determined edge
parameters consistent with the XMM2 dataset. The {\it Chandra} spectrum (CXO1),
which is of lower signal to noise, does
not require an edge, but if it is included, the energy and depth are fully
consistent with the XMM values (see Table 2).

Fig. 2 shows $\chi^2$ confidence contours for the edge energy versus the
optical depth derived from the XMM2 data in comparison with the K edge energy
expected for different ionic species of Fe. The best-fit parameters are
compatible with Fe XVII, with a range from Fe XV to Fe XVIII, indicating
significant ionization of iron. Assuming the absorption cross section of the
Fe XVII K edge ($2.7 \times 10^{-20}$~cm$^2$), 
computed using the analytical fits of Verner \& Yakovlev (1995),
we can determine the column density of
ionized Fe to $N_{Fe} \approx 1.7 \times 10^{19}$~cm$^{-2}$, relatively
independent
of the ionization state of iron. Assuming solar abundances, this would correspond
to an effective hydrogen column density of $N_{H} \approx 6 \times 10^{23}$~cm$^{-2}$,
about a factor of 7 larger than the hydrogen column density
derived by fitting a cold, neutral absorber to the low--energy cutoff
of the spectrum. 

The fit of a neutral absorber in conjunction with a highly ionized Fe-K edge
is, however, physically not meaningful. If Fe is partly ionized, most of the
lighter elements will also be ionized and therefore will 
absorb the X--ray continuum less strongly. Indeed, the detailed treatment of the high
resolution UV/optical spectrum of this source indicates a total neutral
hydrogen column density not much larger than several times $10^{18}$~cm$^{-2}$
(Srianand \& Petitjean, 2000). However, edges of hydrogen- and helium-like 
ions of lower-Z elements (O, Ne, Mg, Si and S)  as well as the Fe L edge
will absorb lower energy photons and can mimic a cold gas absorber (Schartel 
et al. 1997, Komossa 1999). To estimate their effect, we have scaled the 
column density for these elements to the measured $N_{Fe}$, assuming solar
abundance ratios. Including these edges into the model spectrum
produces significantly more low--energy absorption than observed in
the data. Indeed, reducing the K edge depth of lower--Z elements by a
factor of 5, while leaving the depth of the Fe L edge at the predicted
depth of 8.7 gives a very good approximation to the lower--energy
absorption observed in APM 08279+5255. This indicates a high
overabundance of iron with respect to lower-Z elements. 

A more quantitative treatment has to include the photoionization of the BAL
flow and the balance between different ionization states of the elements. 
Therefore we fitted a power law model,
folded through the Galactic $N_H$ and an ionized absorber (model {\em absori}
contained in {\em xspec}, Done et al., 1992). We obtain an excellent fit to the XMM2 data with a
reduced $\chi^2$ of 1.09.  Fig. 3 shows confidence intervals for the 
interesting parameters in the {\it absori} fit. The equivalent hydrogen
column density is in the range 
$N_H = (1.0\pm0.2)  \times 10^{23}$~cm$^{-2}$ 
somewhat higher than the neutral hydrogen column densities
in Table 1, but an iron overabundance of $Fe/O$=2--5 is required to take care of the
observed Fe-K edge in relation to the lower-energy absorption by 
O, Ne, Mg K edges and the Fe L edge. There is a correlation between these
two parameters in the sense that the total Fe column density 
can be determined to $N_{Fe}\approx1.5 \times 10^{19}$~cm$^{-2}$, rather independent
from the Fe overabundance and  
consistent with the value obtained by just fitting an edge (see above).

For a more detailed study, we
have constructed grids of photoionization models using the code {\em
Cloudy} (Ferland 1993),
varying the gas column density $N_{Fe}$ and the ionization parameter $U$
until
we obtain an Fe-K ionization level and optical depth consistent with the
observed edge (see below).
$U$ is defined as $U = {Q \over {4\pi{r}^{2}n_{\rm H}c}}$,
where $Q$ is the rate of ionizing photons above the Lyman limit.
As input continuum we chose a mean AGN  
continuum spectrum,
consisting of piecewise powerlaws from the radio to the gamma-ray region
with, in particular, $\alpha_{\rm uv-x}$=--1.4 in the
EUV and an observed $\Gamma_{\rm x}=-2.0$.  The results of this analysis
are discussed in section 3.1.

\section{Discussion}

For the first time, we have reported the detection of
a highly ionized iron edge in the spectrum of a high-redshift BAL quasar.
This indicates a very high column density, ionized 
absorber in the line of sight. Compared to the warm absorbers
typically observed in Seyfert galaxies, which are dominated
by OVII -- OVIII edges of optical depth around unity
(see Komossa 1999 and references therein), where strong Fe K edges
have not been seen so far, the column density of ionized Fe is
considerably higher here and the absorption is very likely
associated with the highly ionized BAL flow observed in the 
UV spectrum of the source. 

For the quasar 3C 351 it has been shown, that the associated absorption
lines observed in the UV spectrum are most likely 
produced by the same highly ionized
material that is also responsible for the moderate X--ray warm absorber
(Mathur et al., 1994). 
In case of BAL systems, however, 
interpretation is complicated by the uncertainties in 
optical depth measurements of the lines:
the broad, saturated UV absorption
lines from different atomic species do not allow an unambiguous column
density determination (e.g., Hamann 1998).

\subsection{Implications for X--ray Broad Absorption Line Models} 

Murray \& Chiang (1995) constructed a model in which 
the BALs arise in an accretion-disk wind driven by line pressure.  
They predicted that objects with very broad CIV absorption
($>$ 5000 km~s$^{-1}$) 
should also show 
Fe absorption edges in the X-ray regime. 
This is similar 
to what we now observe.  

The high column density inferred for the X-ray absorber in 
APM 08279+5255 is also consistent with
predictions of the unified model by Elvis (2000). In that model,
the UV broad absorption lines occur, when the line of sight to the central
object grazes along the funnel--shaped shell of ionized matter
thought to constitute the BAL outflow.
Using NGC\,5548 
as reference object in constructing his new unified
model,  Elvis (2000) predicted  a column density of a few 10$^{23}$ cm$^{-2}$
for an NGC\,5548-like object viewed
along the funnel. We find $\sim$10$^{23}$ cm$^{-2}$ for 
APM 08279+5255.

The ionization state of the warm absorber is characterized by the
ionization parameter $U$. 
The number rate of ionizing photons, $Q$,  was estimated
from our piecewise powerlaw spectrum with $\alpha_{\rm uv-x}$=--1.4 in the
EUV
and $\Gamma_{\rm x}$ as observed, normalized to the observed X--ray flux.
We then find $Q = 1.2 \times 10^{58} k^{-1}$ s$^{-1}$, $k$ being the lensing magnification 
factor.
Given the best estimate of the ionization parameter 
(log U = 0.4) 
and assuming a density of the absorber of log $n$ = 9.5 -- this
is similar to the value suggested by Elvis (2000) to ensure 
pressure equilibrium with the BELR clouds -- 
we expect a lower limit on the location of the absorber of
$r \approx 2 \times 10^{18} k^{-1/2}$ cm. A more detailed 
discussion of the ionized gas properties and consequences 
for BAL models will be given in a forthcoming paper
(G. Hasinger et al. 2002, in prep.).

\subsection{Relation of X--ray and UV Absorbers and Metal Abundances}

Comparison with UV data immediately implies that the X--ray absorber
must be dust-free, else the UV continuum would be heavily extincted.
The high degree of ionization of the X--ray absorber is consistent
with the UV spectrum which shows strong high-ionization BAL lines
including OVI, but relatively weak neutral H absorption.

The UV absorption line data show that the BAL flow is highly structured
with saturated absorption bands
and thus, contrary to the X--ray absorption edge, cannot determine the total
column density in the flow. 
Based on our best-fit {\em{Cloudy}} model, 
we expect to see UV absorption in e.g.,  HI, CIV, NV, OVI and NeVIII   
with column densities on the order of $N_{\rm HI} \approx 4\,\,10^{16}$cm$^{-2}$, 
$N_{\rm CIV} \approx 10^{16}$cm$^{-2}$,
$N_{\rm NV} \approx 7\,\,10^{17}$cm$^{-2}$, 
$N_{\rm OVI} \approx 6\,\,10^{17}$cm$^{-2}$, and
$N_{\rm NeVIII} \approx 7\,\,10^{17}$cm$^{-2}$.
The higher metal column densities compared to HI can be traced
back to the high degree of ionization of the absorber
and do therefore not necessarily indicate chemical overabundances.
For a detailed comparison with the UV data, {\em simultaneous}
multi-wavelength observations are required. 

No diagnostically valuable, strong Fe lines are present in the UV spectra
of BALs. Thus, iron abundance determinations for BALs 
are still scarce. Fe-K edge X-ray observations therefore 
nicely supplement UV measurements of other metal species,
and allow tests of different scenarios for the origin of the BAL gas. 

We find an overabundance of iron of 2--5 times the solar abundance (see Fig. 3).
Due to the fact that the measured X--ray absorption is produced mainly
by O, Ne, Mg, Si and Fe, this iron overabundance is actually
an estimate of $Fe/O$.  The outflowing
material must therefore have already been processed in a starburst
environment.   
Detailed chemical evolutionary scenarios of the emission-line gas
in quasars (Hamann \& Ferland 1993) predict the iron enrichment
that depends mostly on the lifetime of supernova Type Ia precursors, leading
to an expected delay of $\sim$1 Gyr until $Fe/O$ reaches solar values.
Fe measurements in high-$z$ objects, like APM 08279+5255,
therefore (1) are of profound relevance
for understanding the early star formation history of the universe
and (2) provide important constraints on cosmological models.
Assuming an Fe abundance of APM 08279+5255 of at least solar, we can
place severe constraints on some of the enrichment models of Hamann \&
+Ferland
(1993; their Fig. 1). Furthermore, at the redshift of $z\approx 4$
the age of the universe is a little less than 1 Gyr (for $q_{\rm o}$=0.5),
which compares to the timescale of $\sim$1 Gyr necessary to enrich $Fe/O$ up to
the solar value. Given that we find strong indications
of a supersolar $Fe/O$ abundance, we are beginning to constrain
cosmological models, favoring those which predict larger galaxy
ages at a given $z$. 

In the near future, Fe abundance
measurements of larger samples of high-$z$
quasars may provide another valuable path to measure cosmological parameters
(Hamann \& Ferland 1993).

\acknowledgments

This work is based on observations
obtained with {\it XMM--Newton}, an ESA science mission with instruments and
contributions directly funded by ESA Member States and the USA (NASA).
 We thank the project scientist,
Fred Jansen, for granting directors discretionary time for the
long observation presented here, and the Science
Operation
Center in Villafranca for carrying out the observations efficiently.
On the German side, the {\it XMM--Newton} project is
supported by the Bundesministerium f\"ur Bildung und Forschung/Deutsches
Zentrum f\"ur Luft- und Raumfahrt (BMBF/DLR), the Max-Planck Society
and the Heidenhain-Stiftung. We thank Sarah Gallagher for making available
to us the {\it Chandra} spectrum of APM 08279+5255, and Gary Ferland for 
providing {\em{Cloudy}}. We acknowledge helpful discussions with H. B\"ohringer,
C. Done, P. Petitjean, Th. Boller, W. Brinkmann, D. Porquet and Weimin Yuan. We thank the
referee, Martin Elvis, for very helpful comments.

\clearpage

\begin{figure}
\plotone{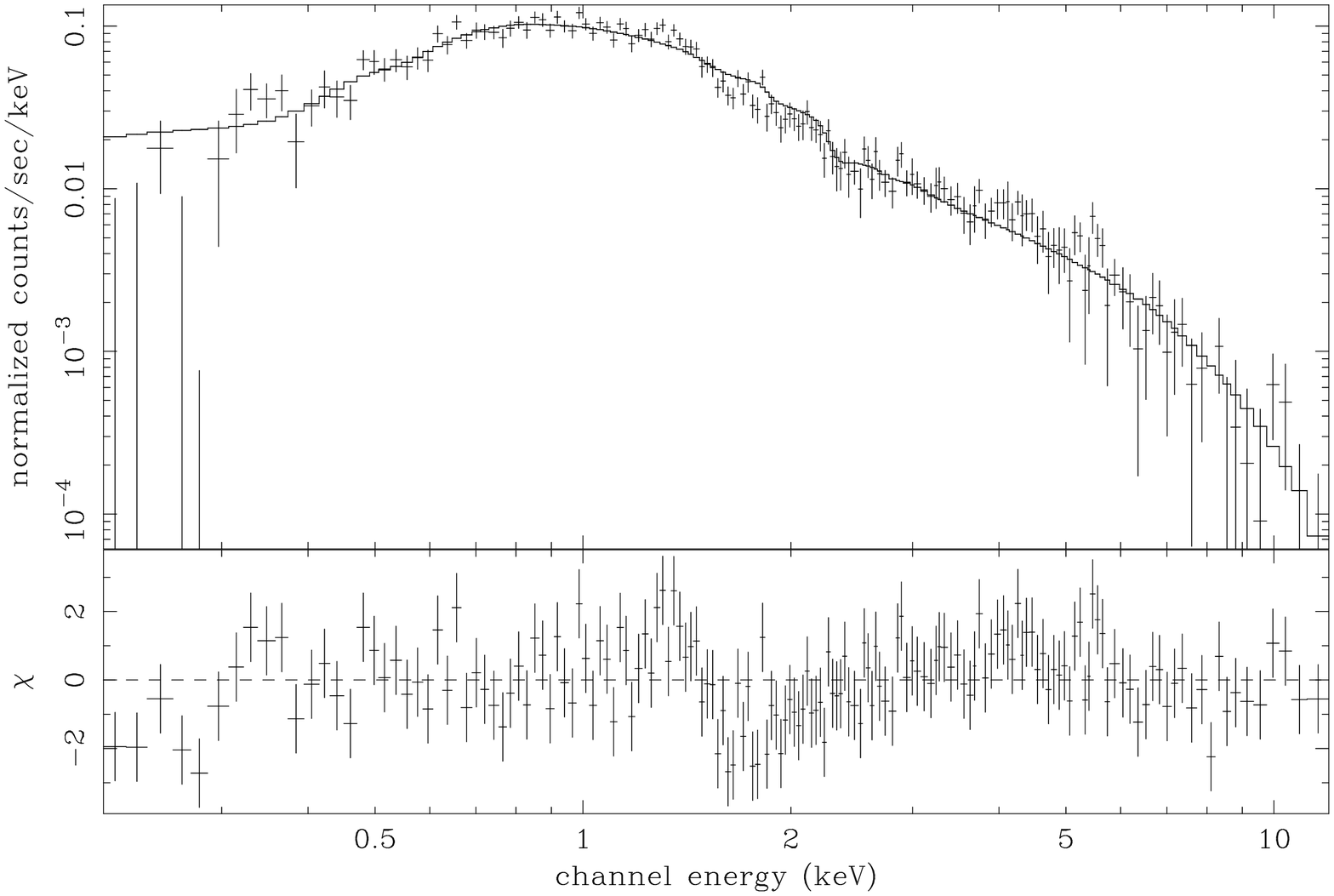}
\caption{ X--ray spectrum of APM 08279+5255 taken with the {\it XMM--Newton} 
pn--CCD camera, fit with a simple power law model absorbed by neutral gas in 
our own Galaxy as well as associated with the source. This fit is 
statistically not acceptable; residuals are shown below the curve.}
\end{figure}

\clearpage
\begin{figure}
\plotone{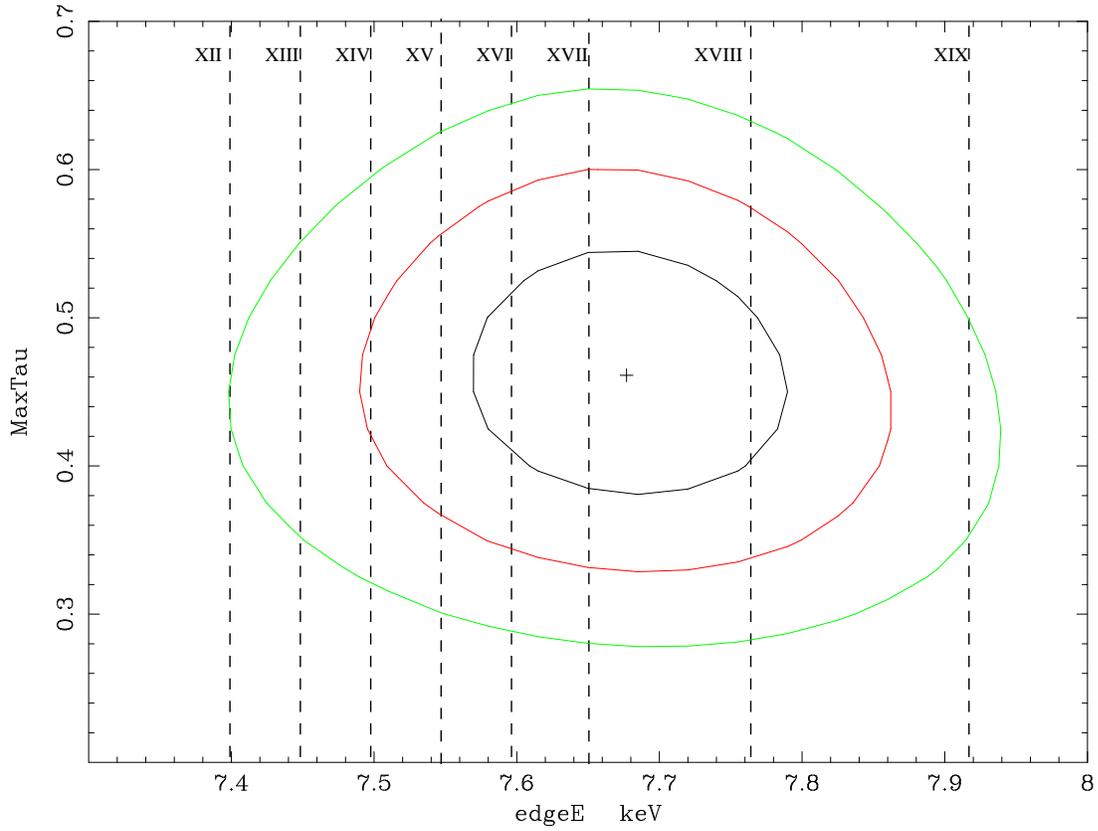}
\caption{Confidence contours of the edge energy versus the optical depth of the
edge. Contours are given for 1, 2 and 3$\sigma$ deviations from the best fit.
The dashed vertical lines refer to the Fe-K edges for the specified
ionic species.}
\end{figure}

\clearpage
\begin{figure}
\plotone{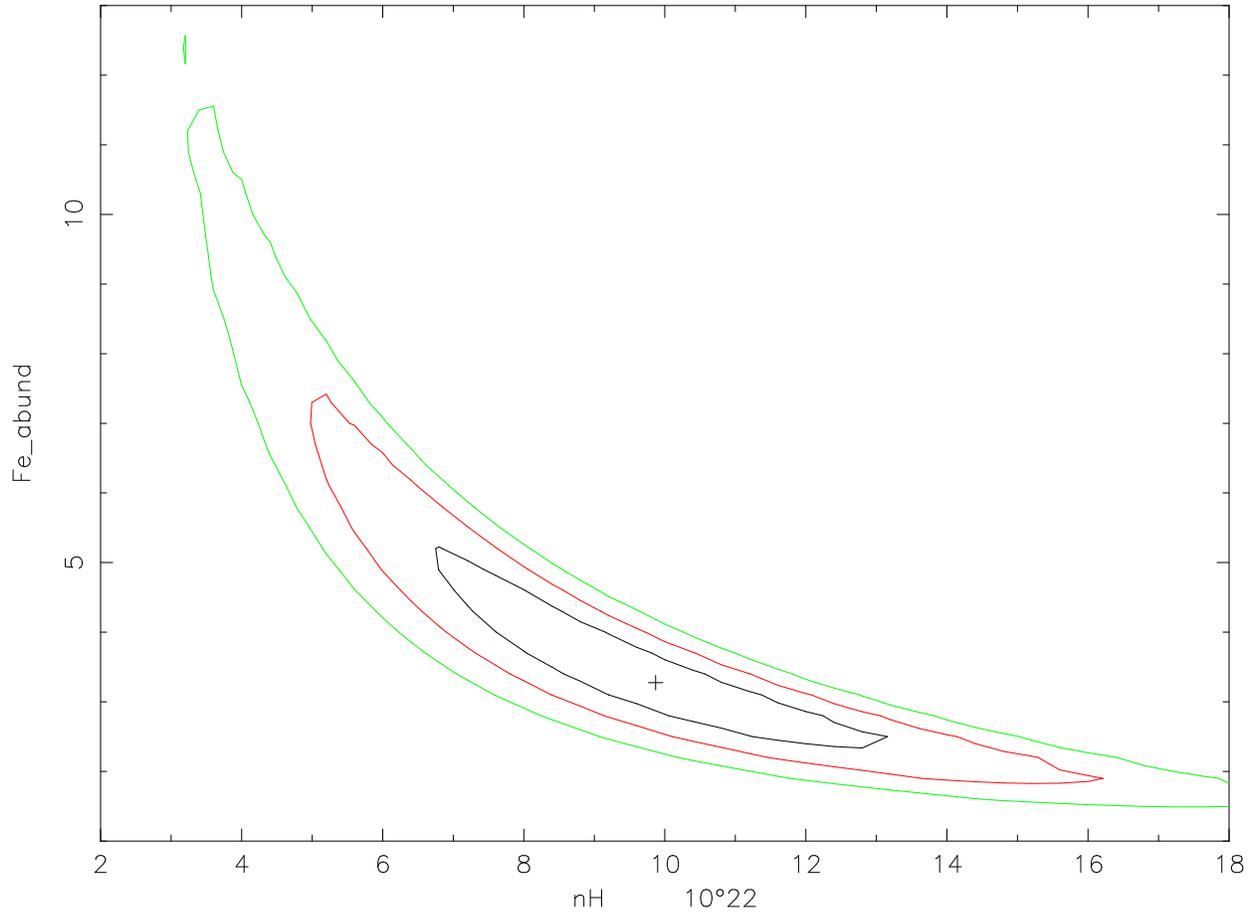}
\caption{Confidence contours of the hydrogen column density $N_H$ versus the iron 
abundance $Fe/O$ in an ionized absorber ({\it absori}) fit.}
\end{figure}

\clearpage
\begin{table}
\begin{center}
\caption{Observing log for APM 08279+5255.}
\begin{tabular}{lcccrl}
\tableline\tableline
ID & Start Date & End Date &  Filter & Exposure & Instrument \\
   &   [UT]   & [UT]     &  & [s]  \\
\tableline
CXO1 & 11.10.2000 16:13 & 11.10.2000 19:24 &        & 9137 & ACIS-I\\
XMM1 & 30.10.2001 02:50 & 30.10.2001 07:35 & medium & 16523 & EPIC\\
XMM2 & 28.04.2002 17:07 & 29.04.2002 21:36 & medium & 102274 & EPIC \\
\tableline
\end{tabular}
\end{center}
\end{table}

\clearpage

\begin{deluxetable}{lcccccccrr}
\tablecaption{Spectral fit results.}
\tablewidth{0pt}
\tabletypesize{\scriptsize}
\tablehead{
\colhead{ID}  & \colhead{$N_H$\tablenotemark{a}} & \colhead{$\Gamma$} & 
\colhead{Norm\tablenotemark{b}} & \colhead{$E_{edge}$} & \colhead{$\tau$} & 
\colhead{$\xi$\tablenotemark{c}} & \colhead{$Fe/H$} &
\colhead{$\chi^2_{red}$} & \colhead{d.o.f.}
}
\startdata
CXO1 & 5.26$\pm$0.77 & 1.78$\pm$0.08 & 1.48$\pm$0.19 & \nodata & \nodata & \nodata & \nodata &  1.05 &  51\\
XMM1 & 6.18$\pm$0.46 & 1.98$\pm$0.04 & 1.30$\pm$0.09 & \nodata & \nodata & \nodata & \nodata &  1.32 & 168\\
XMM2 & 6.92$\pm$0.32 & 2.04$\pm$0.03 & 1.30$\pm$0.05 & \nodata & \nodata & \nodata & \nodata &  1.32 & 365\\
& & & & & & & \\
CXO1 & 5.44$\pm$0.77 & 1.76$\pm$0.08 & 1.55$\pm$0.21 & 7.67$\pm$0.39 & 0.30$\pm$0.16 & \nodata & \nodata & 1.00 &  49\\
XMM1 & 6.67$\pm$0.47 & 1.96$\pm$0.04 & 1.41$\pm$0.10 & 7.62$\pm$0.13 & 0.47$\pm$0.08 & \nodata & \nodata & 1.10 & 165\\
XMM2 & 7.34$\pm$0.34 & 2.01$\pm$0.03 & 1.37$\pm$0.06 & 7.68$\pm$0.10 & 0.46$\pm$0.05 & \nodata & \nodata & 1.09 & 362 \\
& & & & & & & \\
XMM1 & 9.1$\pm$1.4 & 2.00$\pm$0.04 & 1.44$\pm$0.11 & \nodata & \nodata & 29$\pm$23 & 3\tablenotemark{d} & 1.14 & 166   \\
XMM2 & 9.8$\pm$2.3 & 2.03$\pm$0.03 & 1.42$\pm$0.07 & \nodata & \nodata & 47$\pm$12 & 3.3$\pm$0.9 &  1.09 & 362 \\
\enddata
\tablenotetext{a}{absorber at z=3.91 in units of $10^{22}~cm^{-2}$ (for {\it ``absori''}: ionized); all models include Galactic
absorption.}
\tablenotetext{b}{ in $10^{-4}~ph~cm^{-2}~s^{-1}~keV^{-1}$.}
\tablenotetext{c}{Ionization parameter $\xi$ from ionized absorber fit.}
\tablenotetext{d}{fixed in fit.}
\end{deluxetable}

\end{document}